\documentclass{article}
\usepackage{spconf,amsmath,graphicx}
\usepackage{algorithm}
\usepackage{algpseudocode}
\usepackage{amsfonts}
\usepackage{tikz}
\usepackage{tikz-3dplot}
\usepackage{caption}
\usepackage{subcaption}

\DeclareMathOperator*{\argmin}{argmin}

\title{Sensor Array and Camera Fusion via Unbalanced Optimal Transport for 3D Source Localization}
\name{Ilyes JAOUEDI, Gilles CHARDON and José PICHERAL}
\address{Université Paris-Saclay, CNRS, CentraleSupélec, Laboratoire des signaux et systèmes\\ 91190, Gif-sur-Yvette, France}
\begin{document}
\ninept
\maketitle
\begin{abstract}
We address the problem of localizing multiple sources in 3D by combining sensor array measurements with camera observations.
We propose a fusion framework 
extending the covariance matrix fitting
method with an unbalanced optimal transport regularization term that softly aligns sensor array responses with visual priors while allowing flexibility in mass allocation. 
To solve the resulting large-scale problem, we adopt a greedy coordinate descent algorithm that efficiently updates the transport plan. Its computational efficiency makes full 3D localization feasible in practice. The proposed framework is modular and does not rely on labeled data or training, in contrast with deep learning–based fusion approaches. Although validated here on acoustic arrays, the method is general to arbitrary sensor arrays. Experiments on real data show that the proposed approach improves localization accuracy compared to sensor-only baselines.
\end{abstract}
\section{Introduction}
\label{sec:intro}

Source localization, in particular in 3D settings,  is a central problem in array signal processing \cite{526899, pesavento_three_2023}. Classical approaches based solely on sensor arrays---such as beamforming, MUSIC \cite{schmidt_multiple_1986}, CoMET \cite{OTTERSTEN1998185} or Covariance Matrix Fitting (CMF) \cite{yardibi_sparsity_2008}---generally provide poor depth accuracy \cite{HERZOG2022108733,10.1121/10.0035915,10810466}, making it difficult to accurately estimate source positions, and therefore, power \cite{CHARDON2022116544}, in realistic conditions. Cameras, on the other hand, can provide complementary geometric and semantic information by detecting objects in the scene. Several works have investigated fusion of cameras with acoustic arrays \cite{10208791,9423042} or with radar \cite{10225711}, but most of these rely on deep learning models. While effective in some cases, such approaches require large amounts of labeled data and retraining for each new setting (e.g. environment, sensor array and camera geometry, etc.), which limits their generalization and practicality.

The main contribution of this work is a proof of concept for sensor array–camera fusion framework based on optimal transport. The key idea is to integrate camera detections directly into the source estimation process of a sensor array, rather than combining modalities after independent estimation. Specifically, we extend CMF with an unbalanced optimal transport (UOT) regularization. Unlike rigid late-fusion methods that simply restrict the search space to visual detections, UOT acts as an elastic coupling. It softly aligns array responses with visual priors, allowing the camera to provide directional cues without enforcing strict mass conservation, thus making the system robust to calibration errors and partial occlusions. Unlike deep learning–based methods, our approach is modular, does not require training or labeled data, and is computationally efficient since the use of the greedy coordinate descent solver makes the optimization fast, which in turn makes full 3D localization feasible in practice.

In this paper, we focus on acoustic arrays as a concrete test case. Microphone arrays are widely used in source localization and provide a challenging scenario due to noise and the difficulty of estimating source depth. Nevertheless, the proposed framework is not limited to acoustics: since it only requires a sensor array covariance model, the same formulation can be applied to other modalities.

Section 2 outlines related works on source localization, optimal transport and sensor array-camera fusion. Section 3 introduces the proposed fusion framework. Experimental results are given in section 4, and section 5 concludes the paper.

\section{Related Work}

We recall here the building blocks of the proposed framework, i.e. source localization, object detection in images, and optimal transport. Sensor array and camera fusion in acoustic and RADAR settings is also discussed.

\subsection{Acoustic Source Localization Methods}

The major challenges for acoustic source localization is the accuracy of the estimation of the position and powers of the sources, and spatial resolution which is the ability to separate closely spaced sources. Both accuracy and resolution are limited in purely acoustic systems, due to the physical nature of sound wave propagation, in particular the large wavelengths involved, and the limited aperture of microphone arrays. Traditional acoustic localization methods are based on beamforming techniques, relying on time difference of arrival (TDOA) information between microphones. While computationally efficient, these methods suffer from limited spatial resolution.

More advanced techniques, such as CoMET \cite{OTTERSTEN1998185} or Covariance Matrix Fitting (CMF) \cite{yardibi_sparsity_2008}, aim to improve resolution by matching the observed spatial covariance matrix to a modeled response from hypothetical sources. We denote $\mathbf x_\ell$ 
the complex amplitudes of the signals captured on a sensor array at a given frequency of interest and at a time $t_\ell$ (e.g. obtained by a short time Fourier transform). It is modeled as a sum of contributions from $M$ potential sources on a predefined spatial grid \(
    \mathbf x_\ell = \sum_{j=1}^{M} \mathbf g_j s_{j\ell} + \mathbf e_\ell
\)
where $\mathbf g_j$ is the transfer vector for the $j$-th grid point to the sensors, $s_{j\ell}$ is the complex amplitude of that source at time $t_\ell$, and $\mathbf e_\ell$ is an additive noise assumed to be white and uncorrelated. The empirical covariance matrix of the observations over $L$ snapshots is
\[
    \mathbf{\hat{\Sigma}} = \frac{1}{L} \sum_{\ell=1}^{L} \mathbf x_\ell \mathbf  x_\ell^H
\]
For large $L$, the empirical covariance converges to the theoretical covariance matrix 
\(
    \mathbf R = \mathbf G \mathbf Q \mathbf G^H + \sigma^2 \mathbf I
\)
where $\mathbf G = [\mathbf g_1, \ldots, \mathbf g_M]$ is the transfer matrix, $\mathbf Q$ is the diagonal source power matrix with $\mathbf Q = \mathrm{diag}(\mathbf p)$ and $\mathbf p \in \mathbf{R}_+^M$ are the powers of the sources, $\sigma^2$ the noise variance and $\mathbf I$ the identity matrix.

Instead of relying on beamforming, which suffers from poor spatial resolution, or CoMET, which provides higher resolution at the cost of considerable complexity, we adopt the CMF approach as a practical trade-off between accuracy and efficiency.  CMF estimates the source powers $\hat{\mathbf{p}}$ by minimizing the Frobenius norm  $\| \cdot \|_F$ between the modeled and observed covariance matrices, i.e.  

\begin{equation}
    \hat{\mathbf{p}} = \argmin_{\mathbf p \geq 0} \left\| \mathbf G \, \mathrm{diag}(\mathbf p) \, \mathbf G^H - \mathbf{\hat{\Sigma}} \right\|_F^2.
    \label{eq:cmf}
\end{equation}
We note that the proposed optimal transport fusion framework does not rely on the particular properties of the CMF criterion, and that other optimization based array processing can be considered.

As was observed in several numerical and actual experiments, large errors estimation of the distance between the sources and the microphone \cite{HERZOG2022108733,10.1121/10.0035915} (or sensor \cite{10810466}) array are to be expected when the sources are not close to the sensor array. This in turn implies that the power of the source is misestimated, as e.g. an overestimation of the distance will imply an underestimation of the power \cite{CHARDON2022116544}.

\subsection{Object Detection in Images}

Our assumption is that the acoustical sources
are associated to physical objects visible in images captured on a camera (e.g. loudspeakers), and that this visual information can be included in the estimation of the positions of the sources.
Object detection in images is generally based on deep learning methods such as \textit{YOLO (You Only Look Once)}\cite{yolo}. These models can identify and localize objects in an image in the form of bounding boxes.
However, a major challenge lies in estimating the depth of detected objects.
Monocular depth estimation models, such as \textit{MiDaS} \cite{ranftl_towards_2022}, have been developed to overcome this limitation. Still, they remain sensitive to variations in scene conditions and camera viewpoint, which can introduce estimation errors.

In conclusion, image processing can be used to guide the localization of the sources, but cannot estimate all parameters, in particular distance of the sources to the camera, and emitted power.

\subsection{Optimal Transport}

Optimal transport, introduced by Monge and formalized by Kantorovich \cite{peyré2019computational}, provides a mathematical framework to quantify and optimize the cost of moving mass from one distribution to another. In the discrete formulation, we consider two mass distributions: 
\(
\nu_{\mathbf{a}} = \sum_{n=1}^N a_n \delta_{u_n} \quad \text{and} \quad \eta_{\mathbf{b}} = \sum_{m=1}^M b_m \delta_{v_m}
\)
defined by non-negative weights \( \mathbf{a} \in \mathbb{R}^N_+ \), \( \mathbf{b} \in \mathbb{R}^M_+ \), and point sets \( U = (u_1, ..., u_N) \), \( V = (v_1, ..., v_M) \). 

We define a cost matrix \( \mathbf C \in \mathbb{R}^{M\times N} \), where each element \( C_{mn} \) represents the cost of transporting mass from \( u_n \) to \( v_m \).  
The discrete optimal transport problem consists of finding a transport plan \( \mathbf P \in \mathbb{R}^{M \times N}_{+} \) that minimizes the total transport cost
\begin{equation}
\label{ot_discret}
\mathrm{OT}(\nu_{\mathbf{a}},\eta_{\mathbf{b}}) = \min_{\mathbf P \geq 0} \langle \mathbf C, \mathbf P \rangle
\end{equation}
subject to the mass conservation constraints:
$
\mathbf P \mathbf{1}_N = \mathbf{b}$ and $\quad \mathbf P^\top \mathbf{1}_M = \mathbf{a}
$

where $\langle \cdot, \cdot \rangle$ denotes the Euclidean dot product, $\cdot^\top$ matrix transposition, and \( \mathbf{1}_M \), \( \mathbf{1}_N \) are column vectors of dimension \( M \) and \( N \) filled with ones.

In the case of Unbalanced Optimal Transport (UOT) \cite{séjourné2023unbalancedoptimaltransporttheory}, the constraints are relaxed by introducing divergence penalties that account for mass variation. The formulation becomes:
\begin{equation}
\label{uot_discret}
\mathrm{UOT}(\nu_{\mathbf{a}},\eta_{\mathbf{b}})=\min_{\mathbf P \geq 0} \langle \mathbf C, \mathbf P \rangle + \alpha D_1(\mathbf{b}, \mathbf P \mathbf{1}_N) + \beta D_2(\mathbf{a}, \mathbf P^\top \mathbf{1}_M)
\end{equation}
where \( D_{1,2}(\cdot , \cdot) \) are divergence measures penalizing mismatches between the original and transported marginals (e.g., Euclidean norm, Kullback–Leibler divergence), and \( \alpha, \beta \) are regularization parameters. 
As shown in \cite{elvander_multi-marginal_2020}, optimal transport can be applied to sensor fusion to effectively combine information from multiple, homogeneous,  arrays (e.g. multiple microphone arrays).

\subsection{Sensor Array-Camera Fusion}

As outlined in the introduction, several methods have been proposed for the fusion of acoustical data, or in general, sensor array data (in particular RADAR \cite{10225711}), and images. In early fusion, sensor data and image data are processed jointly, e.g. as inputs to a neural network \cite{10208791}. While this approach leverages correlations between the modalities, a limitation of the approach is the necessity to retrain the neural network when e.g. the geometry of the sensor array and the camera is modified.

In late fusion, modalities are first processed independently, e.g. by first extracting a point cloud from the sensor array representing sources or object of interest \cite{8916629}. While this approach may be less sensitive to changes in the geometry (e.g. the point cloud extraction can be model based, exploiting the geometry explicitly) , a late fusion is less capable of leveraging the full potential of the combined data.
In the proposed approach, the estimation of the parameters of the source will exploit the additional prior knowledge obtained by image processing, while ensuring flexibility, in particular with respect to the geometry of the sensor array and the camera. This design reduces the need for costly retraining when hardware changes and improves data localization accuracy. While Deep Learning methods \cite{10208791,9423042,10225711} may offer high performance in fixed settings, they lack this modularity. Our approach provides a training-free alternative that remains robust to layout variations.

\section{CMF--UOT}

We introduce here the proposed  CMF--UOT method , which consists of regularizing CMF with a term based on unbalanced optimal transport, used to align camera detections with the acoustic localization grid.

\subsection{Problem Formulation}

In the proposed application, two heterogeneous modalities are used: acoustics and images.
Image processing will be used to build a proposed distribution of sources, with highly uncertain powers and distances with respect to the camera. An optimal transport will then be used to regularize the CMF problem towards solutions compatible with the partial information given by the image.
The use of unbalanced optimal transport to fuse these modalities is justified by the fact that the camera does not provide reliable estimates of the acoustic power. It is therefore not meaningful to enforce strict mass conservation.
\begin{figure}[t]
    \centering
    \includegraphics[width=\columnwidth]{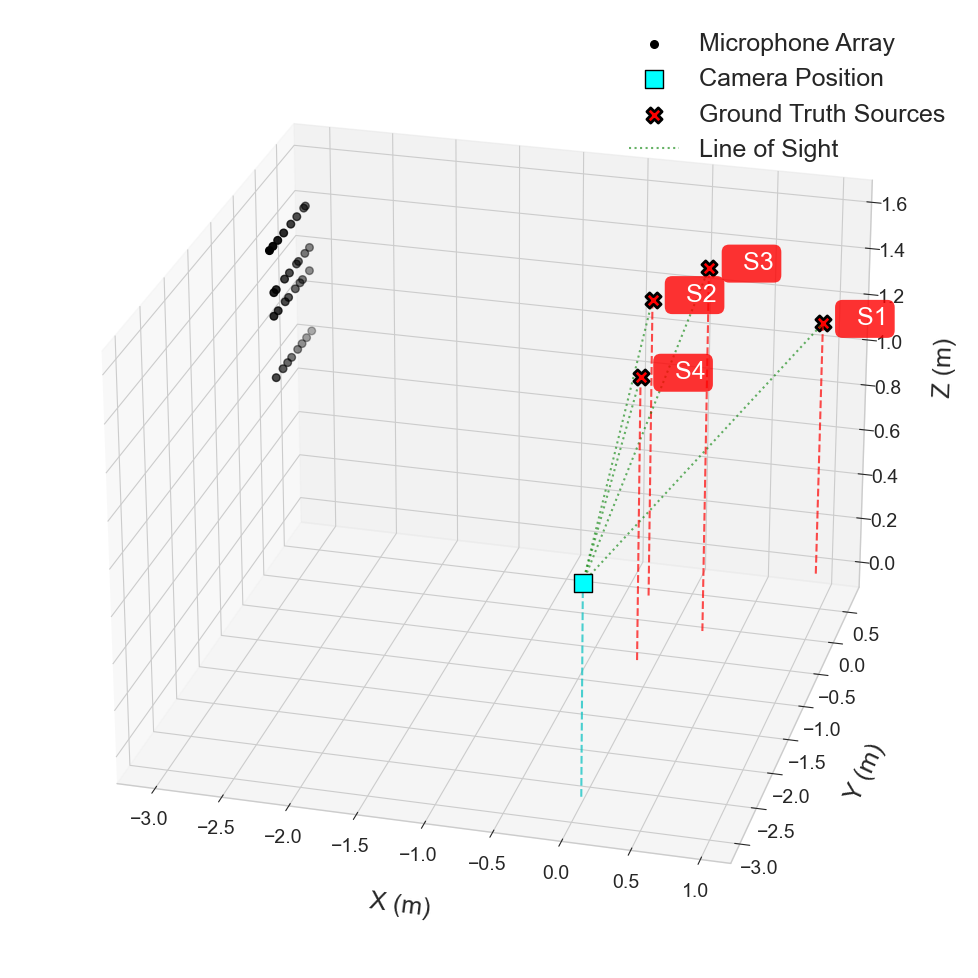}
    \caption{Spatial Configuration}
    \label{fig:teta}
\end{figure}


As illustrated in Figure~\ref{fig:teta}, we consider that the acoustic array and the camera have perpendicular points of view regarding the positions of the sources. 
Object detection from the camera yields a set of $N$ points $u_n = (u^x_n, u^y_n, u^z_n)$ with uniform weights $a_n = 1/N$. The acoustic source distribution to be estimated, denoted by $\eta_{\mathbf{b}}$, is supported on a discretized grid of $M$ points. The weights $b_m \geq 0$ represent the estimated acoustic power at each grid location.

The objective is to estimate the distribution $\eta_{\mathbf{b}}$ using both the acoustic data and the 3D positions extracted from camera detections. To this end, we regularize the CMF problem~\eqref{eq:cmf} by adding a UOT term between the visual detections and the source distribution. This leads to the optimization problem
\begin{equation}
\label{of}
\min_{\mathbf{b} \geq 0} \quad \left\| \mathbf{G} \operatorname{diag}(\mathbf{b}) \mathbf{G}^H - \mathbf{\hat{\Sigma}} \right\|_F^2 + \lambda \, \mathrm{UOT}(\nu_{\mathbf{a}}, \eta_{\mathbf{b}}).
\end{equation}
By substituting the definition of UOT from Equation~\eqref{uot_discret}, choosing $D_1(\cdot, \cdot)$ as zero when arguments are equal and $+\infty$ otherwise, and
a quadratic cost for $D_2$, we obtain the quadratic problem
\[
\min_{\mathbf{b} \geq 0} \left\| \mathbf{G} \operatorname{diag}(\mathbf{b}) \mathbf{G}^H - \mathbf{\hat{\Sigma}} \right\|_F^2 
+ \lambda \min_{\mathbf{P} \geq 0} \left\langle \mathbf{C}, \mathbf{P} \right\rangle + \frac{\beta}{2} \|\mathbf{a} - \mathbf{P}^\top \mathbf{1}_M\|_2^2
\]
subject to the constraint $\mathbf{P} \mathbf{1}_N = \mathbf{b}$.

By incorporating the mass constraint into the formulation and denoting $\mu = \lambda\beta$, the optimization problem can be rewritten as:

\begin{equation}
\label{off}
\min_{\mathbf{P} \geq 0} \left\| \mathbf{G} \operatorname{diag}(\mathbf{P} \mathbf{1}_N) \mathbf{G}^H - \mathbf{\hat{\Sigma}}\right\|_F^2 + \lambda \, \left\langle \mathbf{C}, \mathbf{P} \right\rangle 
+ \frac{\mu}{2} \, \|\mathbf{a} - \mathbf{P}^\top \mathbf{1}_M\|_2^2,
\end{equation}
a quadratic problem with respect to the transport plan $\mathbf P$.





\subsection{Definition of the Cost Matrix}

In this work, we manually select the pixel coordinates of the detections in the camera image. These pixels are then projected into real-world coordinates using reference points located at a fixed depth. As a result, the 3D coordinates $\{u_n = (u^x_n, u^y_n, u^z_n)\}_{n=1}^N$ correspond to the projections of the detections onto the plane defined by the references, rather than exact depth estimates. The sensor array grid is represented by $M$ candidate points $\{v_m = (v^x_m, v^y_m, v^z_m)\}_{m=1}^M$.

To account for the fact that the camera is unable to provide an accurate depth, the transport cost between two points $u_n$ and $v_n$ is based on the (positive) angle $\theta_{nm}$ between their respective line of sights with respect to the camera, i.e.
\begin{equation}
    \mathbf{C}_{nm} = \theta_{nm}
    \end{equation}

This expression measures the arc length on the unit sphere centered at the camera, ensuring that all points lying along the same direction from the camera induce the same cost.

\subsection{Optimization and Problem Resolution}

The optimization problem in Eq.~\eqref{off} involves a large number of variables, as it requires estimating a full transport plan $\mathbf{P} \in \mathbb{R}_+^{M \times N}$ between $M$ grid points and $N$ visual detections. Large optimal transport are frequently tackled by adding an entropy regularization term and using the Sinkhorn algorithm \cite{NIPS2013_af21d0c9}. 
Owing to the particular structure of problem (\ref{off}) as a quadratic problem with non-negativity constraints, we adopt the coordinate descent algorithm with maximal improvement proposed in \cite{gcd}, which updates one entry of $\mathbf{P}$ at a time, choosing at each iteration the coordinate that yields the maximal decrease in the objective function.

\section{Experimental results}
\begin{figure}[H]
    \centering
    \includegraphics[width=\linewidth]{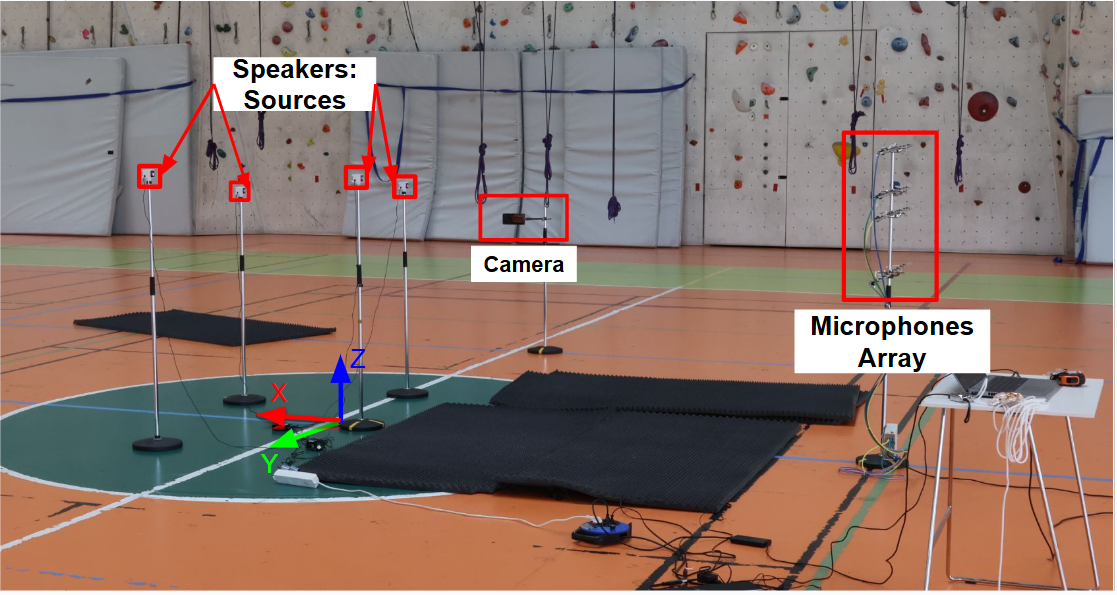}
    \caption{Experimental Setup}
    \label{fig:setup}
\end{figure}

To validate the proposed approach, we collected a dataset in the gymnasium of the Bouygues building at CentraleSupélec. The experimental setup, illustrated in Figure~\ref{fig:setup}, includes four omnidirectional broadband loudspeakers (Visaton BF32, 150~Hz–20~kHz), a smartphone used as camera, and a microphone array composed of 32 MEMS microphones. The microphones are distributed irregularly along four horizontal lines, covering an area of approximately 1~m in width and 0.6~m in height.

\begin{figure}[t]
    \centering
    \begin{subfigure}[b]{0.9\columnwidth}
        \centering
        \includegraphics[width=\textwidth]{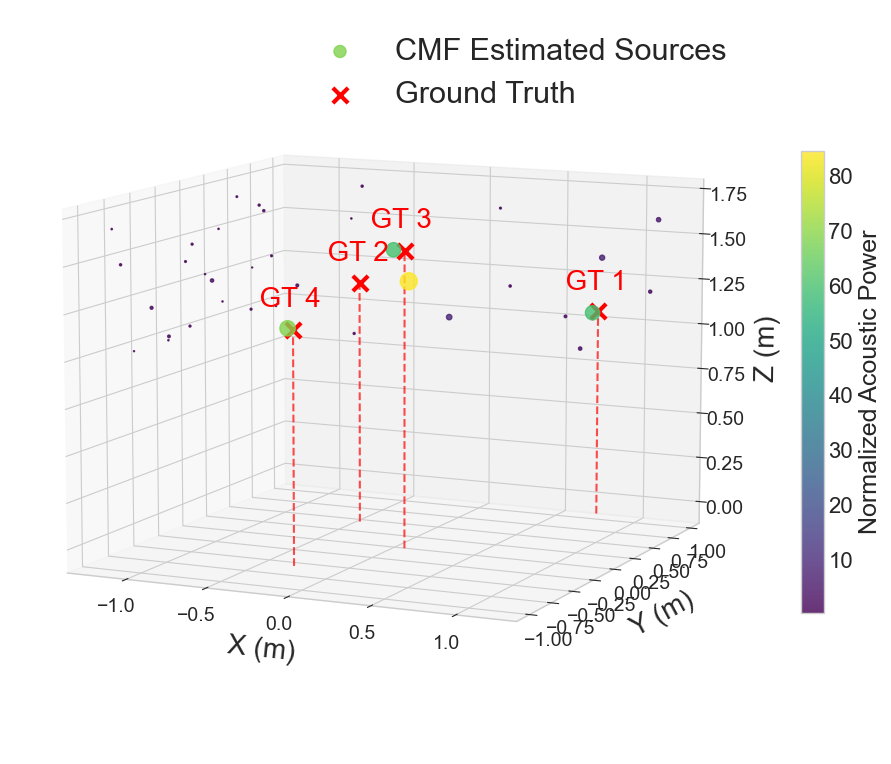}
        \caption{Full point cloud with CMF.}
        \label{fig:3dcloud_cmf}
    \end{subfigure}
    \hfill
    \begin{subfigure}[b]{0.9\columnwidth}
        \centering
        \includegraphics[width=\textwidth]{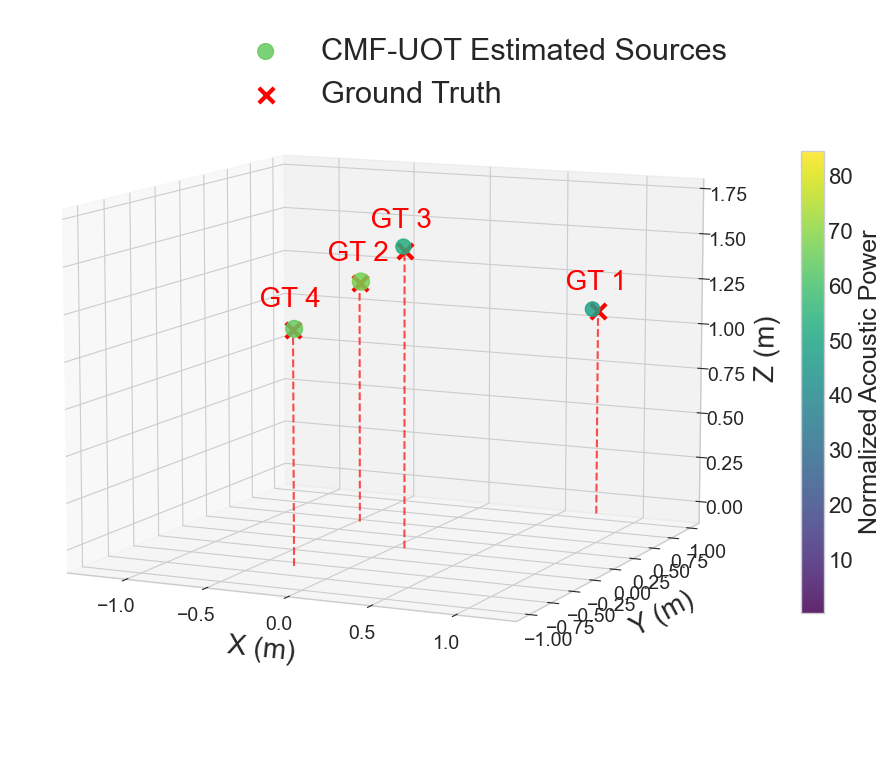}
        \caption{Full point cloud with CMF--UOT.}
        \label{fig:3dcloud_cmf_uot}
    \end{subfigure}
    \caption{Estimated source distributions on the 3D grid.}
    \label{fig:3dcloud}
\end{figure}

\begin{figure}[t]
    \centering
    \includegraphics[width=\columnwidth]{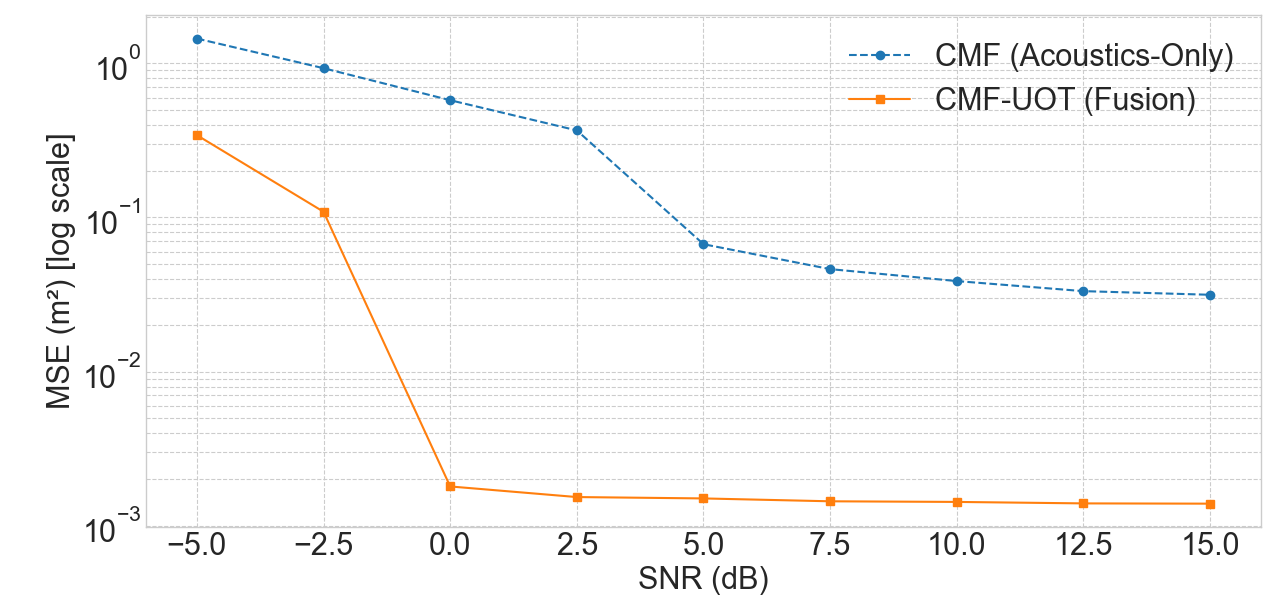}
    \caption{Mean squared localization error versus SNR. }
    \label{fig:snr}
\end{figure}

\begin{figure}[t]
    \centering
    \includegraphics[width=\columnwidth]{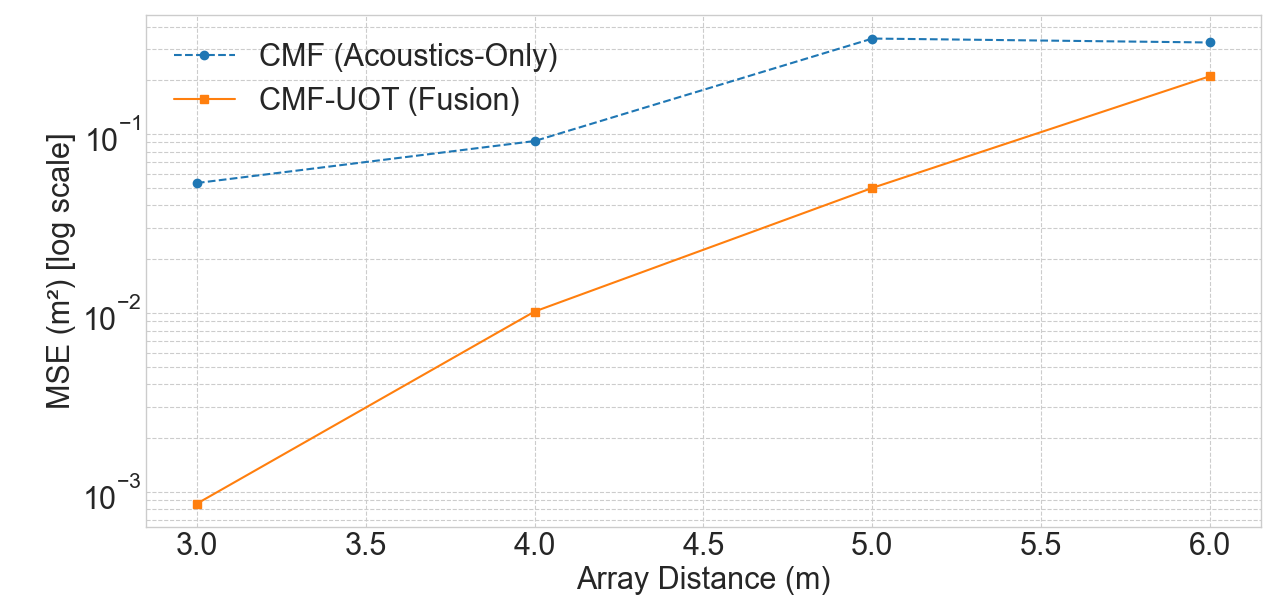}
    \caption{Mean squared localization error as a function of source distance from the array (fixed SNR = 10~dB).}
    \label{fig:distance}
\end{figure}

The excitation frequency of the sources is 4~kHz, and $L = 513$ snapshots were recorded. The search area is discretized into a grid of $M = 
4.2\times 10^5$ candidate points covering a volume of 2.4~m $ \times $ 2~m $ \times $ 0.7~m = 3.36~m$^3$  with step 0.02~m. Regularization parameters were empirically chosen as $\lambda = 2 \times 10^{2}$ and $\mu = 5 \times 10^{-4}$. Qualitatively, $\lambda$ controls the strength of the visual prior; a very low $\lambda$
 reverts to standard CMF, while an excessive $\lambda$
 forces the acoustic power to concentrate strictly on visual rays. The parameter $\mu$
 controls the 'unbalanced' relaxation, allowing the total estimated mass to deviate from the visual detections.
Figure~\ref{fig:3dcloud} shows the estimated source distributions obtained with CMF (using the Lawson-Hanson algorithm \cite{CHARDON2021116208}) and with the proposed CMF--UOT method. The full point clouds illustrate the localization results across the search grid compared to  the ground-truth sources. CMF produces diffuse estimates with spurious artifacts, while CMF--UOT, on the right, yields sharper and more spatially accurate localization. The improvement is particularly visible along the $x$-axis, where the array alone struggles to estimate the depth of source~2.
Localization accuracy is further evaluated using the mean squared error (MSE) between estimated and ground-truth source positions. Each point in the following plots is averaged over 100 independent runs. We assume the initial recordings to be noise-free and synthetically add Gaussian noise to reach the desired SNR level. Figure~\ref{fig:snr} reports the MSE as a function of SNR. The proposed CMF--UOT consistently improves performance over CMF, particularly at low SNR where the array-only method fails to resolve sources. At high SNR, CMF exhibits a larger residual error than CMF--UOT due to one misestimated source position (source~2), as illustrated in Figure~\ref{fig:3dcloud}. An error floor remains visible for both methods, reflecting limitations of the experimental setup: loudspeakers are not perfectly point-like (3.2~cm diameter), ground-truth positions are annotated manually, and variability in the measurements introduces additional uncertainty.
To further analyze robustness, we varied the distance of the sources from the array, from $3$~m up to $6$~m, at a fixed SNR of $10$~dB. Figure~\ref{fig:distance} shows that CMF--UOT consistently achieves lower error than CMF across all tested distances. The improvement is maintained at larger distances, where the angular resolution of the array alone becomes insufficient.

All experiments were performed on a computer equipped with an Intel Core i7 processor (2.30~GHz) and 16~GB of RAM. On this hardware, one run of CMF--UOT for 20 iterations requires 2.45~seconds, which demonstrates the computational efficiency of the proposed greedy coordinate descent solver and confirms the feasibility of 3D localization in practice.

\section{Conclusion}
We presented a proof of concept for a sensor array--camera fusion framework based on optimal transport. The proposed approach is computationally efficient and highly modular: the CMF criterion can be replaced by alternative optimization based array processing methods, and the visual front-end can rely on any suitable neural network architecture for object detection. Experiments on real acoustic data confirmed sharper and more accurate localization compared to array-only baselines. Future work will aim to extend this proof-of-concept beyond the current static scene assumption. Additionally, more advanced mass-transport constraints could better handle severe occlusions where acoustic sources are visually obstructed. 
Also, we plan to investigate automatic tuning of the regularization parameters $\lambda$ and $\mu$ to adapt them according to the estimated noise level of the measurements and the confidence of camera detections, for example using the apparent size of the detected objects. This would further improve robustness and adaptability to different sensing conditions.

\newpage

\section{ACKNOWLEDGMENT}
This research work benefited
from the support of the Chair "Massive and Heterogenous Data processing for Smart Vehicules" operated by CentraleSupélec
and sponsored by Forvia.

\bibliographystyle{IEEEbib}
\bibliography{strings,refs}

\begin{thebibliography}{10}

\bibitem{526899}
H.~{Krim} and M.~{Viberg},
\newblock ``Two decades of array signal processing research: the parametric
  approach,''
\newblock {\em IEEE Signal Processing Magazine}, vol. 13, no. 4, pp. 67--94,
  July 1996.

\bibitem{pesavento_three_2023}
Marius Pesavento, Minh Trinh-Hoang, and Mats Viberg,
\newblock ``Three {More} {Decades} in {Array} {Signal} {Processing} {Research}:
  {An} optimization and structure exploitation perspective,''
\newblock {\em IEEE Signal Processing Magazine}, vol. 40, no. 4, pp. 92--106,
  June 2023,
\newblock Conference Name: IEEE Signal Processing Magazine.

\bibitem{schmidt_multiple_1986}
R.~Schmidt,
\newblock ``Multiple emitter location and signal parameter estimation,''
\newblock {\em IEEE Transactions on Antennas and Propagation}, vol. 34, no. 3,
  pp. 276--280, Mar. 1986.

\bibitem{OTTERSTEN1998185}
B~Ottersten, P~Stoica, and R~Roy,
\newblock ``Covariance matching estimation techniques for array signal
  processing applications,''
\newblock {\em Digital Signal Processing}, vol. 8, no. 3, pp. 185--210, 1998.

\bibitem{yardibi_sparsity_2008}
Tarik Yardibi, Jian Li, Petre Stoica, and Louis~N. Cattafesta,
\newblock ``Sparsity constrained deconvolution approaches for acoustic source
  mapping,''
\newblock {\em The Journal of the Acoustical Society of America}, vol. 123, no.
  5, pp. 2631--2642, May 2008.

\bibitem{HERZOG2022108733}
Adrian Herzog and Emanuël~A.P. Habets,
\newblock ``Distance estimation in the spherical harmonic domain using the
  spherical wave model,''
\newblock {\em Applied Acoustics}, vol. 193, pp. 108733, 2022.

\bibitem{10.1121/10.0035915}
Anique Altena, Mirjam Snellen, Salil Luesutthiviboon, Guido de~Croon, and Mark
  Voskuijl,
\newblock ``Frequency band analysis and comparison of localisation techniques
  for drones using microphone array measurements,''
\newblock {\em JASA Express Letters}, vol. 5, no. 2, pp. 024802, 02 2025.

\bibitem{10810466}
Jiaxiong Fang, Hua Chen, Wei Liu, Songjie Yang, Chau Yuen, and Hing~Cheung So,
\newblock ``Three-dimensional localization of mixed near-field and far-field
  sources based on a unified exact propagation model,''
\newblock {\em IEEE Transactions on Signal Processing}, vol. 73, pp. 245--258,
  2025.

\bibitem{CHARDON2022116544}
Gilles Chardon,
\newblock ``Theoretical analysis of beamforming steering vector formulations
  for acoustic source localization,''
\newblock {\em Journal of Sound and Vibration}, vol. 517, pp. 116544, 2022.

\bibitem{10208791}
Wenru Zheng, Ryota Yoshihashi, Rei Kawakami, Ikuro Sato, and Asako Kanezaki,
\newblock ``Multi event localization by audio-visual fusion with
  omnidirectional camera and microphone array,''
\newblock in {\em 2023 IEEE/CVF Conference on Computer Vision and Pattern
  Recognition Workshops (CVPRW)}, 2023, pp. 2566--2574.

\bibitem{9423042}
Bin Duan, Hao Tang, Wei Wang, Ziliang Zong, Guowei Yang, and Yan Yan,
\newblock ``Audio-visual event localization via recursive fusion by joint
  co-attention,''
\newblock in {\em 2021 IEEE Winter Conference on Applications of Computer
  Vision (WACV)}, 2021, pp. 4012--4021.

\bibitem{10225711}
Shanliang Yao, Runwei Guan, Xiaoyu Huang, Zhuoxiao Li, Xiangyu Sha, Yong Yue,
  Eng~Gee Lim, Hyungjoon Seo, Ka~Lok Man, Xiaohui Zhu, and Yutao Yue,
\newblock ``Radar-camera fusion for object detection and semantic segmentation
  in autonomous driving: A comprehensive review,''
\newblock {\em IEEE Transactions on Intelligent Vehicles}, vol. 9, no. 1, pp.
  2094--2128, 2024.

\bibitem{yolo}
Joseph Redmon, Santosh~Kumar Divvala, Ross~B. Girshick, and Ali Farhadi,
\newblock ``You only look once: Unified, real-time object detection,''
\newblock in {\em 2016 {IEEE} Conference on Computer Vision and Pattern
  Recognition, {CVPR} 2016, Las Vegas, NV, USA, June 27-30, 2016}. 2016, pp.
  779--788, {IEEE} Computer Society.

\bibitem{ranftl_towards_2022}
René Ranftl, Katrin Lasinger, David Hafner, Konrad Schindler, and Vladlen
  Koltun,
\newblock ``Towards {Robust} {Monocular} {Depth} {Estimation}: {Mixing}
  {Datasets} for {Zero}-{Shot} {Cross}-{Dataset} {Transfer},''
\newblock {\em IEEE Transactions on Pattern Analysis and Machine Intelligence},
  vol. 44, no. 3, pp. 1623--1637, Mar. 2022.

\bibitem{peyré2019computational}
G.~Peyr{\'e} and M.~Cuturi,
\newblock {\em Computational Optimal Transport: With Applications to Data
  Science},
\newblock Foundations and trends in machine learning. Now Publishers, 2019.

\bibitem{séjourné2023unbalancedoptimaltransporttheory}
Thibault Séjourné, Gabriel Peyré, and François-Xavier Vialard,
\newblock ``Chapter 12 - unbalanced optimal transport, from theory to
  numerics,''
\newblock in {\em Numerical Control: Part B}, Emmanuel Trélat and Enrique
  Zuazua, Eds., vol.~24 of {\em Handbook of Numerical Analysis}, pp. 407--471.
  Elsevier, 2023.

\bibitem{elvander_multi-marginal_2020}
Filip Elvander, Isabel Haasler, Andreas Jakobsson, and Johan Karlsson,
\newblock ``Multi-marginal optimal transport using partial information with
  applications in robust localization and sensor fusion,''
\newblock {\em Signal Processing}, vol. 171, pp. 107474, June 2020.

\bibitem{8916629}
Felix Nobis, Maximilian Geisslinger, Markus Weber, Johannes Betz, and Markus
  Lienkamp,
\newblock ``A deep learning-based radar and camera sensor fusion architecture
  for object detection,''
\newblock in {\em 2019 Sensor Data Fusion: Trends, Solutions, Applications
  (SDF)}, 2019, pp. 1--7.

\bibitem{NIPS2013_af21d0c9}
Marco Cuturi,
\newblock ``Sinkhorn distances: Lightspeed computation of optimal transport,''
\newblock in {\em Advances in Neural Information Processing Systems}, C.J.
  Burges, L.~Bottou, M.~Welling, Z.~Ghahramani, and K.Q. Weinberger, Eds. 2013,
  vol.~26, Curran Associates, Inc.

\bibitem{gcd}
Julie Nutini, Mark Schmidt, Issam~H. Laradji, Michael Friedlander, and Hoyt
  Koepke,
\newblock ``Coordinate descent converges faster with the gauss-southwell rule
  than random selection,''
\newblock in {\em Proceedings of the 32nd International Conference on
  International Conference on Machine Learning - Volume 37}. 2015, ICML'15, p.
  1632–1641, JMLR.org.

\bibitem{CHARDON2021116208}
Gilles Chardon, José Picheral, and François Ollivier,
\newblock ``Theoretical analysis of the damas algorithm and efficient
  implementation of the covariance matrix fitting method for large-scale
  problems,''
\newblock {\em Journal of Sound and Vibration}, vol. 508, pp. 116208, 2021.

\end{thebibliography}

\end{document}